\documentclass[pdflatex,sn-nature]{sn-jnl}

\usepackage{graphicx}

\usepackage{graphicx}%
\usepackage{multirow}%
\usepackage{amsmath,amssymb,amsfonts}%
\usepackage{amsthm}%
\usepackage{mathrsfs}%
\usepackage[title]{appendix}%
\usepackage{xcolor}%
\usepackage{textcomp}%
\usepackage{manyfoot}%
\usepackage{booktabs}%
\usepackage{algorithm}%
\usepackage{algorithmicx}%
\usepackage{algpseudocode}%
\usepackage{listings}%

\def\Msun{\,M$_{\odot}$}
\def\arcmin{^{'}}
\def\arcsec{^{''}}

\def\Hyp{\textit{Hyper}}
\def\Hypy{\textit{Hyper-py}}

\def\rms{\textit{r.m.s.}}

\begin{document}

\title[Article Title]{The fragmentation properties of massive star-forming regions in 30Dor-10 at 2000 au resolution}


\author*[1]{\fnm{Alessio} \sur{Traficante}}\email{alessio.traficante@inaf.it}

\author[2,3]{\fnm{Mar\'ia J.} \sur{Jim\'enez-Donaire}}

\author[4,5]{\fnm{Remy} \sur{Indebetouw}}

\author[6]{\fnm{Tony} \sur{Wong}}

\author[1,7]{\fnm{Alice} \sur{Nucara}}

\author[8,9,10,11]{\fnm{Ralf} \sur{Klessen}}

\author[12]{\fnm{Patrick} \sur{Hennebelle}}

\author[12]{\fnm{Ugo} \sur{Lebreuilly}}

\author[1]{\fnm{Chiara} \sur{Mininni}}

\author[1]{\fnm{Sergio} \sur{Molinari}}

\author[2,13,14]{\fnm{Elena} \sur{Sabbi}}

\author[15, 1]{\fnm{Juan} \sur{Soler}}

\affil*[1]{\orgdiv{INAF-Istituto di Astrofisica e Planetologia Spaziali (INAF-IAPS)}, \orgaddress{\street{Via Fosso del Cavaliere 100}, \city{Rome}, \postcode{00133}, \state{Italy}}}

\affil[2]{\orgdiv{AURA for ESA, Space Telescope Science Institute}, \orgaddress{\street{3700 San Martin Drive}, \city{Baltimore}, \state{MD 21218}, \country{USA}}}

\affil[3]{\orgdiv{Observatorio Astron\'omico Nacional (IGN)}, \orgaddress{\street{C/Alfonso XII 3}, \city{Madrid}, \postcode{28014}, \country{Spain}}}

\affil[4]{\orgdiv{Astronomy Department, University of Virginia}, \orgaddress{\street{530 McCormick Rd}, \city{Charlottesville}, \state{VA 22904}, \country{USA}}}

\affil[5]{\orgdiv{National Radio Astronomy Observatory}, \orgaddress{\street{520 Edgemont Road}, \city{Charlottesville}, \state{VA 22903}, \country{USA}}}

\affil[6]{\orgdiv{Department of Astronomy, University of Illinois}, \orgaddress{\city{Urbana}, \state{IL 61801}, \country{USA}}}

\affil[7]{\orgdiv{Dipartimento di Fisica, Università di Roma Tor Vergata}, \orgaddress{\street{Via della Ricerca Scientifica 1}, \postcode{I-00133}, \city{Rome}, \country{Italy}}}

\affil[8]{\orgdiv{Universitat Heidelberg, Zentrum f\"ur Astronomie, Institut f\"ur Theoretische Astrophysik}, \orgaddress{\street{Albert-Ueberle-Str. 2}, \postcode{69120}, \city{Heidelberg}, \state{Germany}}}

\affil[9]{\orgdiv{Universitat Heidelberg, Interdisziplin\"ares Zentrum f\"ur Wissenschaftliches Rechnen}, \orgaddress{\street{Im Neuenheimer Feld 205}, \postcode{69120}, \city{Heidelberg}, \country{Germany}}}

\affil[10]{\orgdiv{Harvard-Smithsonian Center for Astrophysics}, \orgaddress{\street{60 Garden Street}, \city{Cambridge}, \state{MA, 02138}, \country{USA}}}

\affil[11]{\orgdiv{Elizabeth S. and Richard M. Cashin Fellow at the Radcliffe Institute for Advanced Studies at Harvard University}, \orgaddress{\street{10 Garden Street}, \city{Cambridge}, \state{MA, 02138}, \country{USA}}}

\affil[12]{\orgdiv{AIM, CEA, CNRS, Universit\'e Paris-Saclay, Universit\'e Paris Diderot}, \orgaddress{\city{Sorbonne Paris Cit\'e, Gif-sur-Yvette}}, \postcode{91191} \country{France}}

\affil[13]{\orgdiv{Gemini Observatory/NSF’s NOIRLab}, \orgaddress{\street{950 North Cherry Avenue}, \city{Tucson}, \state{AZ, 85719}, \country{USA}}}

\affil[14]{\orgdiv{Steward Observatory, University of Arizona}, \orgaddress{\street{933 North Cherry Avenue}, \city{Tucson}, \state{AZ, 85721}, \country{USA}}}

\affil[15]{\orgdiv{Department of Astrophysics, University of Vienna}, \orgaddress{\street{Turkenschanzstrasse 17}, \postcode{1180}, \city{Vienna}, \country{Austria}}}


\abstract{



The fragmentation properties of parsec-scales clumps play a fundamental role in shaping the dense gas condensations known as cores, the immediate progenitor of stars. The distribution of core masses, the so-called core mass function, is the precursor of the stellar initial mass function, which governs the distribution of stellar masses and, consequently, the evolution of galaxies. The stellar initial mass function is often described by a typical Salpeter-like slope, although deviations toward more top-heavy distributions have been reported in extreme environments, raising questions about its universality and about the physical connection between the two mass functions. To date, there are no observational constraints on the core mass function and its link to the initial mass function beyond the Milky Way.

Here we present a study of the fragmentation properties and the measurement of the core mass function in an external galaxy, focusing on the 30Dor-10 region in the Large Magellanic Cloud, using high resolution observations that probe spatial scales down to 2000 au. Robust statistical analysis demonstrates that the core mass function is consistent with a Salpeter-like slope and suggests that variations in the stellar mass distribution arise from evolutionary processes rather than from initial fragmentation.
}




\maketitle

\section*{Introduction}\label{sec:introduction}
The formation of massive stars is a multi-scale process, from tens of parsec-long molecular clouds to high-mass, pc-scale clumps \citep[][]{Zinnecker07, Vazquez-Semadeni25}. Within these clumps, turbulence, gravity, and feedback drive the formation of dense structures at various scales, from about $0.1$ pc objects \citep{Csengeri17} to compact “cores” with radii $500\leq R\leq 4000$ au \citep{Beuther18, Svoboda19, Anderson21, Traficante23, Xu24, Morii24, Ishihara24, Louvet24, Coletta25}. These cores, organized in proto-clusters, are the progenitors of stellar populations \citep[][]{Sanhueza19}. The Initial Mass Function (IMF) describes the mass distribution of these stars, shaping galaxy evolution through stellar feedback. The high-mass IMF tail follows a power-law slope of the form $d\mathrm{N}/d\mathrm{Log(m)}\propto m^{-\Gamma}$ with $\Gamma=-1.35$ \citep{Salpeter55}, long considered “universal” \citep[][]{Kroupa01, Chabrier02}, based on the observation that it appears consistently in a wide range of Galactic environments. The notion of universality has been interpreted as evidence that the dominant physical processes regulating high-mass star formation, such as turbulent fragmentation, radiative feedback, and accretion dynamics, may operate in a broadly self-similar fashion across different star-forming regions. Observational evidence for IMF variations has however also been reported \citep[][]{Bastian10} (but see also \citep{Kroupa13}), for example in particular environments such as open clusters \citep[][]{Prisinzano03}, where both top-heavy and bottom-heavy slopes have been inferred, and in more extreme environments such as the Galactic Center \citep[][]{Hosek19}, massive starburst clusters \citep[][]{Espinoza09} and external galaxies such as the Large Magellanic Cloud \citep[LMC,][]{Banerjee12, Schneider18}, where top-heavy slopes have been observed. In such environments, the high-mass IMF slope can flatten significantly compared to the canonical Salpeter value of $\Gamma = -1.35$, with observed slopes ranging from $\Gamma = -1.0$ to as flat as $\Gamma = -0.7$, depending on the properties of the regions such as mass and metallicity \citep[][]{Marks12, Schneider18}. The effects of radiative feedback and magnetic fields also play a central role in shaping the IMF \citep{Hennebelle20, Hennebelle22}. In particular higher gas temperatures and turbulence levels in starburst regions or galactic centers can raise the Jeans mass, favoring the formation of more massive stars and resulting in a top-heavy IMF \citep{Marks12}. As a consequence, low metallicity environments, where radiative cooling is less efficient, should favor the formation of more massive stars \citep{Marks12}. This is coherent to what is observed in the LMC \citep{Schneider18}, where the metallicity is of the order of $-0.38$ dex on average \citep{Choudhury16} (although different numerical simulations indicate that metallicity may play a secondary role compared to the ambient gas density \citep{Marks12,Tanvir24, Gjergo26}).

Understanding the role of the environment in shaping the IMF is therefore a fundamental aspect of star formation studies \citep[][]{Offner14, Bastian10}. Given the hierarchical and multi-scale nature of star formation, one way to probe the origin of the IMF is by examining its immediate precursor: the mass distribution of dense cores within molecular clouds, i.e. the core mass function (CMF). The CMF can be defined across a broad range of physical scales, from about $0.1$ pc objects both in our Galaxy and nearby galaxies \citep[][]{Konyves15, Indebetouw20}, and down to $1000-2000$ au compact cores in our own Milky Way, at distances up to 7.5 kpc \citep[][]{Louvet24, Coletta25}. At 0.1 pc scales the CMF primarily traces the fragmentation of the parental cloud and the global mass distribution of the environment. At smaller scales, around $1000–5000$ au, the CMF plays a dual role: it acts as the immediate progenitor of the stellar IMF, while also preserving the imprint of large-scale fragmentation and the environmental conditions—such as density, temperature, and metallicity—that may influence the final stellar mass distribution \citep{Hennebelle08, Hennebelle20}.

Such connection between IMF and CMF, however, is not straightforward. Theoretical models suggest that the relation between CMF and IMF depends critically on the dynamical evolution of the cores and the timescales over which they collapse. If both distributions are assumed to be time-invariant, then a given IMF slope would require a CMF with a similar slope (or a shallower one, due to the faster collapse of the more massive cores \citep{Clark07}). Conversely, in time-evolving scenarios, environmental changes may alter the CMF and the measured one at a given time may not correspond directly to the actual IMF \citep{Nunez-Castineyra24}. In particular, the CMF observed in relatively young star-forming regions may represent an earlier evolutionary stage, which could gradually evolve toward a shape that resembles the final IMF as cores accrete mass, merge, or become dispersed \citep{Nucara25}.

Observational studies in the Milky Way have shown that the CMF typically resembles the IMF in shape, often following a log-normal distribution at low masses and a power-law tail at higher masses with a slope close to that of the Salpeter IMF \citep[][]{Andre10, Konyves15}. This apparent similarity has led to the hypothesis that the IMF is largely inherited from the CMF, modulated by a core-to-star efficiency factor \citep[][]{Clark07, Clark21, Offner14, Nunez-Castineyra24}. However, this may not be the case in more extreme environments: environmental factors such as particularly low metallicity, exceptionally high level of turbulence or strong radiation fields can induce temporal variations in the CMF slope, which might only at later stages resemble the IMF slope.

While IMF studies in Galactic and extragalactic environments are extensive \citep{Kroupa01, Chabrier02, Banerjee12, Hopkins18, Schneider18}, CMF studies, in particular across massive and extreme environments, remain challenging due to the rarity and distance of young proto-clusters and the deeply embedded nature of the cores. High-resolution observations have only become feasible with NOEMA and ALMA, enabling systematic studies of star-forming regions in the Milky Way down to $1000-2000$ au resolution. Large programs such as ALMAGAL \citep{Molinari25} and ALMA-IMF \citep{Motte22} have provided statistically significant samples, investigating fragmentation properties and the CMF evolution in massive Galactic clumps \citep{Coletta25}. Emerging results suggest a “clump-fed” scenario \citep{Vazquez-Semadeni19, Anderson21, Coletta25}, where the CMF slope may initially trace the mass distribution of the parent clouds and evolves from a Salpeter-like slope—or even steeper—towards a shallower, top-heavy distribution dominated by massive cores \citep{Motte18a, Pouteau23, Nony23, Armante24, Louvet24, Coletta25, Morii26}.

These findings suggest that in extreme Galactic environments, the IMF may consistently exhibit a top-heavy distribution. In relatively young regions, the CMF is still evolving, potentially toward a more top-heavy shape, therefore a direct correspondence with the final IMF may not yet be observable.

Studies of the CMF across different environments remain fundamental, as they offer the most direct observational insight into fragmentation processes and environmental conditions that shape the formation of cores. They are also key to assessing whether a direct link exists with the IMF, or, if a top-heavy IMF is observed while the CMF is not, whether this may indicate an evolutionary trend in which the CMF progressively shifts toward a top-heavy distribution over time.

In this context, the Large Magellanic Cloud (LMC), with its sub-solar metallicity and intense star formation activity in regions such as the 30 Doradus complex, provides a unique laboratory to investigate how environmental conditions may influence the CMF and, by extension, the resulting IMF. The 30Dor-10 complex in particular \citep{Johansson98} is among the best-characterized extragalactic star-forming regions. It lies at about 15 pc away in projection from R136, a massive star cluster driving intense feedback \citep{Sabbi12, Rahner18, Dominguez23}. While this feedback affects diffuse molecular gas \citep{Indebetouw13} and cloud envelopes \citep{Indebetouw24}, its impact on dense clumps and cores remains unclear \citep{Wong22}. Its CMF derived from $^{12}$CO at about 0.1 pc scales follows a Salpeter-like slope \citep{Indebetouw20}, whereas the IMF in the region is much shallower and cannot be reconciled with a Salpeter-like distribution \citep{Schneider18}. This makes 30Dor-10 a unique laboratory for investigating the multi-scale CMF and how the local environment may affect the origin of the IMF in an environment vastly different from the Milky Way \citep{Brunetti19}. Until now, however, such studies at the necessary resolution of a few thousand au have been prohibited by the distance of the LMC, approximately 50 kpc \citep{Pietrzynski19}, i.e. about 10 times larger then the average distance of the most relevant star-forming complexes in our Galaxy \citep{Motte22}. 

In this work we present a measurement of the global core mass function (CMF) in an external galaxy, achieving a spatial resolution of about 2000 au in massive clumps within the 30Dor-10 region of the LMC. This is made possible by the unique capabilities of ALMA, which combine high angular resolution with exceptional sensitivity, enabling the characterization of the CMF beyond the Milky Way.

\section*{Results}\label{sec:analysis}
\textbf{Star-forming cores in the 30Dor-10 region at 2000 au resolution}. Our data were taken as part of the ALMA project 2022.1.00917.S (PI: A. Traficante) in July 2023. The observations were centered on three massive clumps first identified in the ALMA Cycle 7 project 2019.1.00843.S (PI: R. Indebetouw) \citep{Indebetouw13} with a resolution of about $0.5$\,pc. Here we employ their nomenclature, and refer to them as Clump 4, Clump 52 and Clump 72. These clumps are among the most massive in the region, with CO masses of about 2200 \Msun, 8500 \Msun\ and 7400 \Msun\ within a beam-deconvolved radius of about 0.3 pc, 0.5 pc and 0.5 pc for clump 4, 52 and 72, respectively \citep[][]{Indebetouw13}. 

\begin{figure*}
   \centering
   \includegraphics[width=12cm]{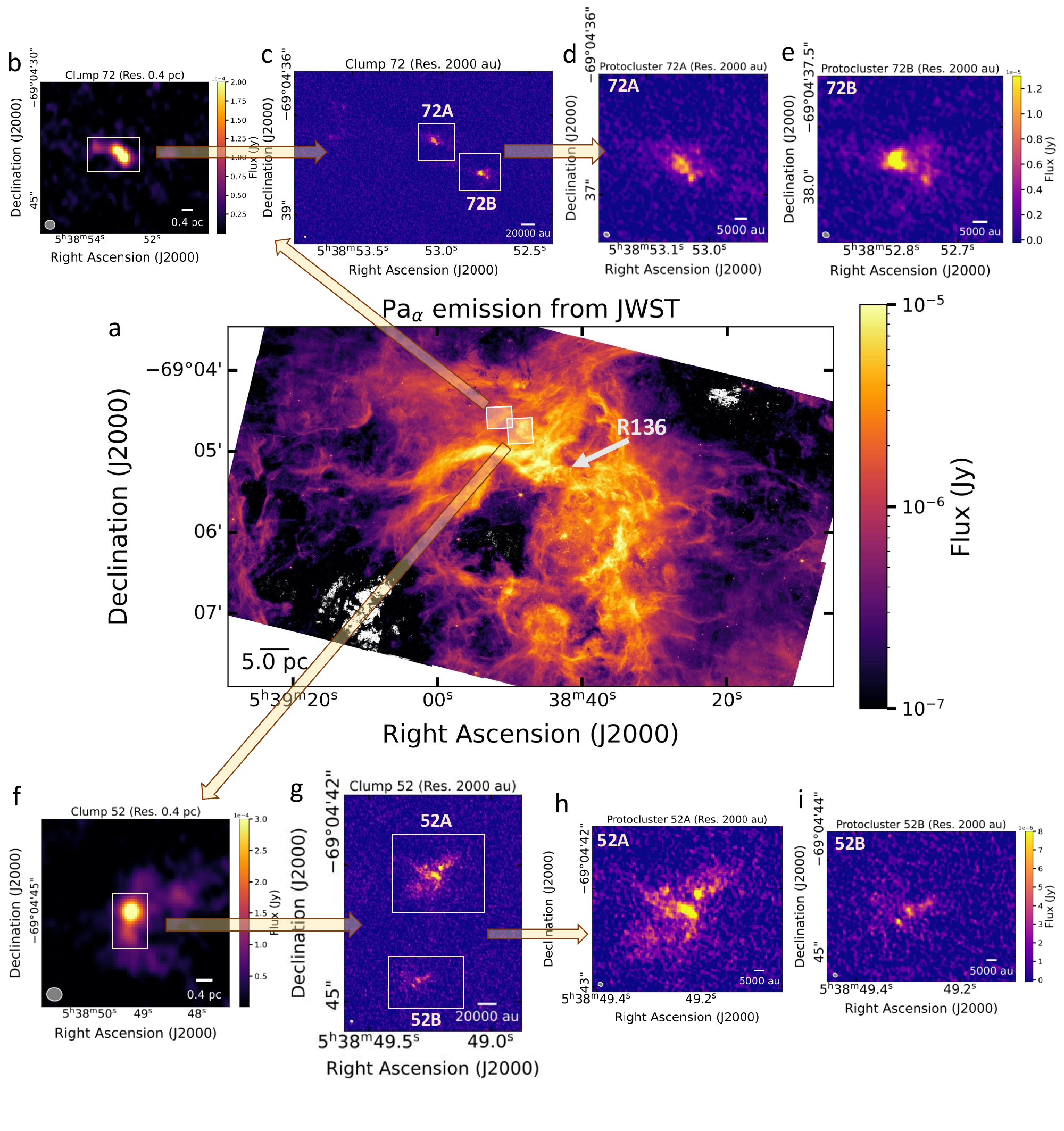}
    \caption{\textbf{Image of the 30Dor-10 region as seen by JWST and ALMA.} \textbf{a} Pa$_{\alpha}$ emission at 1.875\,$\mu$m of the 30Dor-10 region as seen by JWST (taken from the MAST archive). The white boxes are the regions of our two clumps observed with ALMA. The white arrow shows the position of the R136 massive star cluster. \textbf{b} zoomed-in view of clump 72 showing the ALMA 1.3 mm continuum archival data at 0.4 pc resolution (ALMA Cycle 7 project 2019.1.00843.S, \cite{Wong22}). The white box is the region observed with our ALMA data at 2000 au resolution. \textbf{c} 2000 au resolution view of clump 72 from our ALMA 1.3mm continuum data. \textbf{d} Zoom-in of the 2000 au resolution view of proto-cluster 72A. \textbf{e} Zoom-in of the 2000 au resolution view of proto-cluster 72B. \textbf{f} zoomed-in view of clump 52 showing the ALMA 1.3 mm continuum archival data at 0.4 pc resolution (ALMA Cycle 7 project 2019.1.00843.S, \cite{Wong22}). The white box is the region observed with our ALMA data at 2000 au resolution. \textbf{g} 2000 au resolution view of clump 52 from our ALMA 1.3mm continuum data. \textbf{h} Zoom-in of the 2000 au resolution view of proto-cluster 52A. \textbf{i} Zoom-in of the 2000 au resolution view of proto-cluster 52B.}
    \label{fig:JWST_ALMA}
\end{figure*}

Figure \ref{fig:JWST_ALMA} presents, at its center, the $1.875\ \mu$m JWST image of the 30Dor-10 region. The top and bottom panels display the ALMA 1.3 mm continuum images of clump 72 and clump 52, respectively, at a resolution of about 0.35 pc \citep{Wong22} and with our ALMA data at about 2000 au (0.01 pc) resolution. The most striking result is that both clumps are resolved into dense proto-clusters, each comprising multiple compact cores. For the subsequent analysis, we subdivide the clumps into four proto-clusters: 72A and 72B for clump 72, and 52A and 52B for clump 52, corresponding to the northern and southern substructures, respectively.

The cores in our 1.3 mm continuum ALMA data have been identified with a commonly used algorithm in Galactic astronomy, \Hyp\ \citep{Traficante15a}, used also to extract the flux properties. For this work we have used the newly developed version of the code, \textit{Hyper-py} (publicly available at \texttt{https://github.com/Alessio-Traficante/hyper-py} \citep{Traficante25}). Given the intrinsic limitations of interferometric imaging, the identification of compact cores is inevitably subject to uncertainties, potentially propagating into all results based on extracted core samples. While these limitations cannot be eliminated, their impact can be quantified and mitigated through careful validation. Accordingly, we conducted an extensive validation campaign designed to rigorously characterize the cores identified with \Hypy\ and to define a reliable final working catalogue. As detailed in Methods, these tests include the identification of potential imaging artifacts, the run of dedicated simulations to reproduce the complex ALMA background using the Rosetta Stone (RS) pipeline \citep{Lebreuilly25, Tung25, Nucara25}, and the quantification of potential free–free contamination to the measured 1.3 mm ALMA fluxes. For this last test, we utilized ancillary data from the Hubble Space Telescope (HST) and the James Webb Space Telescope (JWST). The HST data consist of continuum-subtracted observations of the 30Dor-10 region used to trace H$_{\alpha}$ line emission with the F658N filter (derived from public available HST image of the region towards the Hubble Tarantula Treasury Project, \texttt{https://archive.stsci.edu/hlsp/http}). The JWST data were retrieved from the MAST archive, and we specifically used the F187N filter (1.875\ $\mu$m) to trace the Pa$_{\alpha}$ emission line. As detailed in the Methods section, our analysis focuses on clump 52 and clump 72, as clump 4 is excluded due to significant free-free contamination ($\geq 30\%$ of the mm flux).

\begin{figure*}
   \centering
   \includegraphics[width=12.0cm]{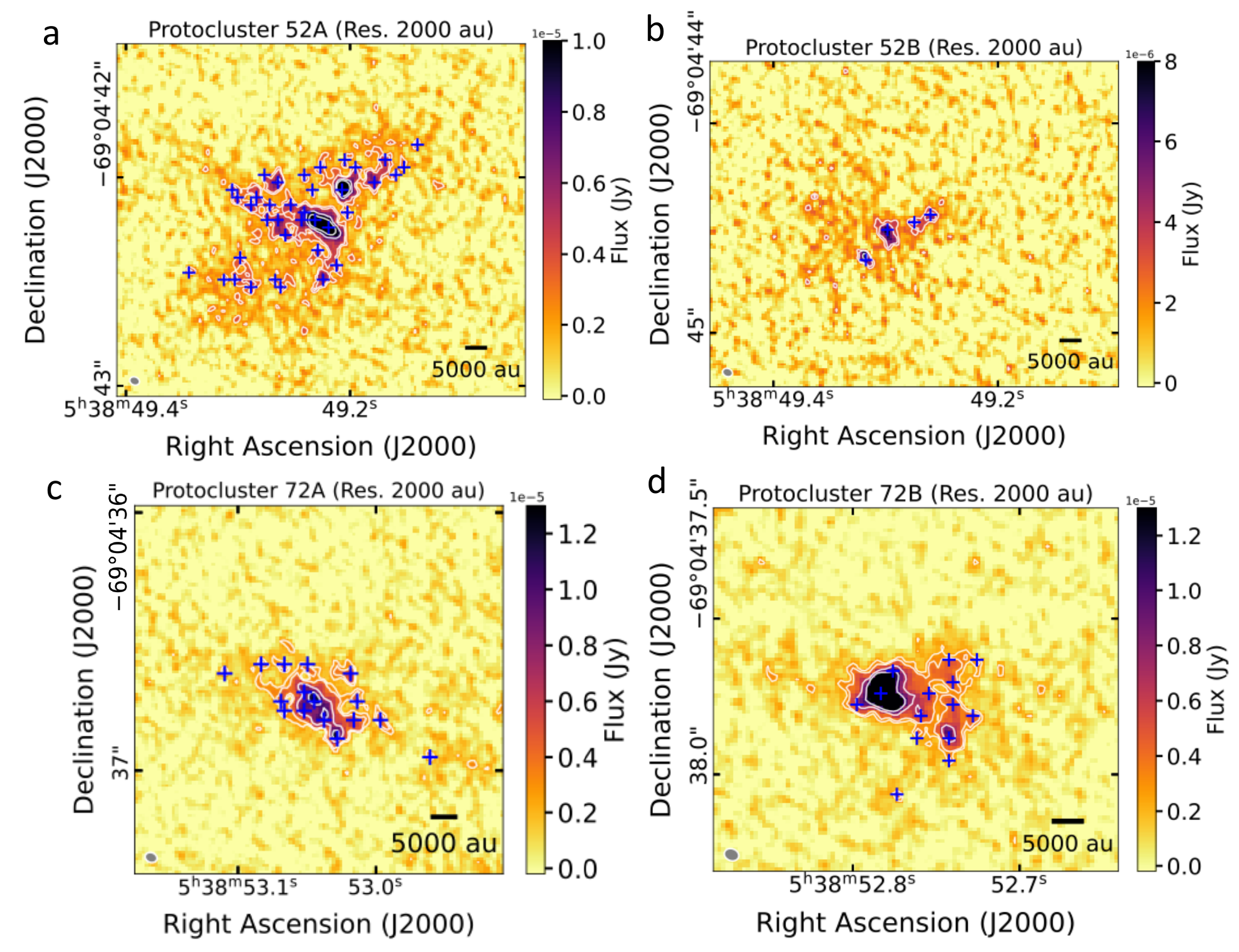} 
    \caption{\textbf{ALMA view of our four proto-clusters at 2000 au with \Hypy\ core identification.} \textbf{a} proto-cluster 52A. \textbf{b} proto-cluster 52B. \textbf{c} proto-cluster 72A. \textbf{d} proto-cluster 72B. In each panel, the blue crosses mark the locations of the \Hypy\ cores identified in each field, and the contours are 3$\sigma$, 4$\sigma$, 8$\sigma$ and 12$\sigma$ level contours with respect to the map \rms.}
    \label{fig:ALMA_cores}
\end{figure*}

In the following analysis we focus on the 71 cores identified within these 4 proto-clusters, as detailed in Methods. These sources are compact, slightly elongated objects (aspect ratio $1-1.5$) with average radii $2150\leq R\leq 3250$ au (estimated as the geometric mean of the two FWHMs derived from each Gaussian fit, and adopted as the semi-axes of the elliptical aperture). Figure \ref{fig:ALMA_cores} presents a more detailed view at 2000 au resolution of our four proto-clusters with highlighted the 71 sources selected for this work.

The 1.3 mm flux, $F_{\mathrm{1.3mm}}$, has been converted into mass using the standard formula for greybody emission, assuming that the dust is optically thin in our regime \citep{Sanhueza19, Svoboda19, Coletta25}:

\begin{equation}\label{eq:mass_from_flux}
    M_{\mathrm{c}} = \frac{R_\mathrm{gd}\,F_{\mathrm{1.3mm}}\,d^{2}}{\kappa_{\mathrm{1.3mm}}\,B_{\mathrm{1.3mm}}(T)}
\end{equation}

\noindent in which $B_{\mathrm{1.3mm}}(T)$ is the Planck blackbody function at 1.3 mm. The quantities $d$ (the distance of our sources), $R_{gd}$ (the gas-to-dust mass ratio), $\kappa_{1.3mm}$ (the dust absorption coefficient) and $T$ (the dust temperature) all contribute to the uncertainties in our analysis.

Different assumptions for $d$, $R_{gd}$ or $\kappa_\nu$ affect the absolute core masses, but they do not impact the CMF slope (under the assumption that they are approximately uniform across the sample), which is the key result of this study. As discussed in details in Methods, we assume $d=50$ kpc, $R_{gd}=500$ and $\kappa_{1.3\mathrm{mm}} = 1.16$ cm$^2$ g$^{-1}$ \citep{Pietrzynski19, Brunetti19}.

The most critical parameter in Equation \ref{eq:mass_from_flux} for our analysis is the dust temperature $T$, which remains challenging to constrain for cores at a spatial scale of 2000 au cores. Determining the temperature from emission lines at these scales and densities is particularly difficult, as these tracers can be contaminated by feedback processes such as outflows \citep{Indebetouw13} or by overlapping gas components at different temperatures \citep{Motte18a, Anderson21, Traficante23, Pouteau22, Louvet24, Coletta25}. Additionally, they can rapidly become optically thick, further complicating temperature estimates \citep{Cunningham23}.

Given the lack of a optimal method for determining $T$, we adopt a reference temperature of 36.7 K, based on range of temperatures averaged across 30Dor-10, N159W, and N159E \citep{Brunetti19}. However, applying a uniform temperature to all cores is likely an oversimplification, making it essential to account for temperature uncertainties.

To address this, we refined methods used in ALMAGAL \citep{Molinari25, Coletta25} and ALMA-IMF \citep{Pouteau22, Louvet24}, performing 5000 Monte Carlo realizations where each core mass was estimated by randomly assigning T from a uniform distribution of 20–80 K \citep{Anderson21, Louvet24, Coletta25}. We further tested the results using different Monte Carlo catalogs (detailed in the Methods section) and found consistent results. Therefore, we present our findings based on the first set of 5000 Monte Carlo realizations.

With the adopted assumptions, core masses in our sample range from approximately 0.4 to 113\Msun, with a mean value of 13.8\Msun\ at the reference temperature. The most massive core, located in the proto-cluster 52A, has a mass of $113.5^{+124.8}_{-65.7}$\Msun, with uncertainties reflecting the adopted temperature range ($T=20$–$80$ K). These values are consistent with the most massive cores found in the ALMAGAL \citep[345\Msun;][]{Coletta25} and ALMA-IMF \citep[150\Msun;][]{Louvet24} surveys, noting that the absolute values depend sensitively on temperature assumptions.

Assuming that the IMF is directly inherited from the actual CMF, a typical core-to-star efficiency of about $50\%$ and that the core forms a single star, the stellar mass expected from our most massive core is $m_{\rm max} = 56.7^{+62.4}_{-32.7}$\Msun. We have compared this estimate with the theoretical expectations for clustered star formation derived from a canonical IMF \citep{Kroupa13}. Such a relation is discussed in the framework of massive ATLASGAL clumps with embedded HII regions, where high-mass stars are found to form in clumps with similar physical properties to those observed in our sample \citep{Zhou24} (although we do not expect a huge contamination from HII regions in our sample, as discussed in Methods). We can use \citep[see Eq. 4 in ][]{Zhou24}:

\begin{equation}\label{eq:mass_star_mass_clumps}
M_{\mathrm{ecl}} \approx \frac{5.37}{\dfrac{0.77}{m_{\mathrm{max}}^{1.3}} - 0.001} 
- \frac{3.33}{m_{\mathrm{max}}^{0.3} \left( \dfrac{0.77}{m_{\mathrm{max}}^{1.3}} - 0.001 \right)} \quad \mathrm{M}_{\odot}
\end{equation}

\noindent where $M_{\mathrm{ecl}}$ is the final mass in stars of the natal clump and $m_{\mathrm{max}}=56.7^{+62.4}_{-32.7}$ \Msun\ is the mass of the most massive star formed in our proto-clusters. We derive a corresponding embedded stellar cluster mass of $M_{\rm ecl} = 1438^{+7020}_{-1078}$,\Msun. This is in good agreement with the available mass reservoir: proto-cluster 52A (the one where we observe the most massive core) is embedded in a CO clump of total mass about 8500 \Msun\ \citep{Indebetouw13}, which is expected to be divided between the two proto-clusters discussed later (52A and 52B) and the diffuse medium. Alternatively, we can estimate the expected most massive stellar mass $m_{\mathrm{max}}$ of the cluster assuming a canonical IMF from the stellar mass of the natal proto-cluster M$_{\mathrm{ecl}}$ \citep{Yan23}. If for example we assume M$_{\mathrm{ecl}}$ of the order $4000-7000$ \Msun, $m_{\mathrm{max}}$ is of the order of $m_{\mathrm{max}} = 60-80$ \Msun\ \citep{Yan23}, in very good agreement with our estimation of the mass of the most massive star that is forming in our proto-clusters. Although a detailed mass decomposition of the original CO clump is beyond the scope of this work, these analysis suggest that the observed core mass distribution is broadly compatible with the formation of a canonical IMF \citep{Kroupa13, Yan23}, rather then a shallower one.

We have used the masses of the 71 cores to build our CMF. To facilitate a statistical analysis of the core mass distribution, we construct a combined, global CMF by merging the core populations extracted from the four proto-clusters identified in the 30Dor-10 region and regardless of their evolutionary stage (i.e., starless, prestellar, or protostellar). Importantly, the four proto-clusters analyzed here are not drawn from unrelated regions, but are physically embedded within only two distinct CO clumps separated by a projected distance of about 5 pc \citep{Indebetouw13}, and thus are likely part of a coherent cloud complex. An analysis of their radio continuum properties (see Methods) also suggests that these systems are at relatively similar evolutionary stages, or at least not significantly different, supporting a meaningful comparison of their internal core populations. In this context, the approximation of combining the core populations of the four proto-clusters into a single CMF is justified as a mean to increase the statistical significance of our analysis. We nevertheless emphasize that this CMF represents the superposition of four compact sub-regions, rather than a single monolithic star-forming event. This is analogous to the IMF derived for this region (excluding, as we also do, RCW136 \citep{Schneider18}), which includes the most massive stars out of the 452 stars distributed across the broader 30 Doradus region, and therefore likely reflects a composite stellar population. The IMF of a region formed by multiple sub-clusters is not expected to match the IMF of any one of its constituents, due to non-linear scaling relations and the self-regulation of high-mass star formation \citep{Kroupa24}. A similar caveat applies to our CMF: while it allows us to probe the global behavior of the dense core population in this environment, it does not necessarily reflect the CMF of any single clump, and should be interpreted accordingly. 

Furthermore, given the uncertainties affecting core mass identification and estimates, no single power-law exponent for the CMF can be easily defined. Although multiple diagnostics indicate that the source extraction is reliable, the intrinsic uncertainties associated with core identification in interferometric images motivated a conservative assessment of potential spurious detections among the lowest-contrast cores, as detailed in Methods. These tests confirm the stability of our results: the CMF slope distribution remains statistically unchanged even when the sample size is progressively reduced. We therefore consider the full sample of 71 cores and use it to construct a CMF for each of the 5000 Monte Carlo catalogs. 

In the following, we derive the CMFs and the slopes of the power-law tails in the 30Dor-10 region with all the assumptions just discussed.

\begin{figure*}
\centering
\includegraphics[width=12.5cm]{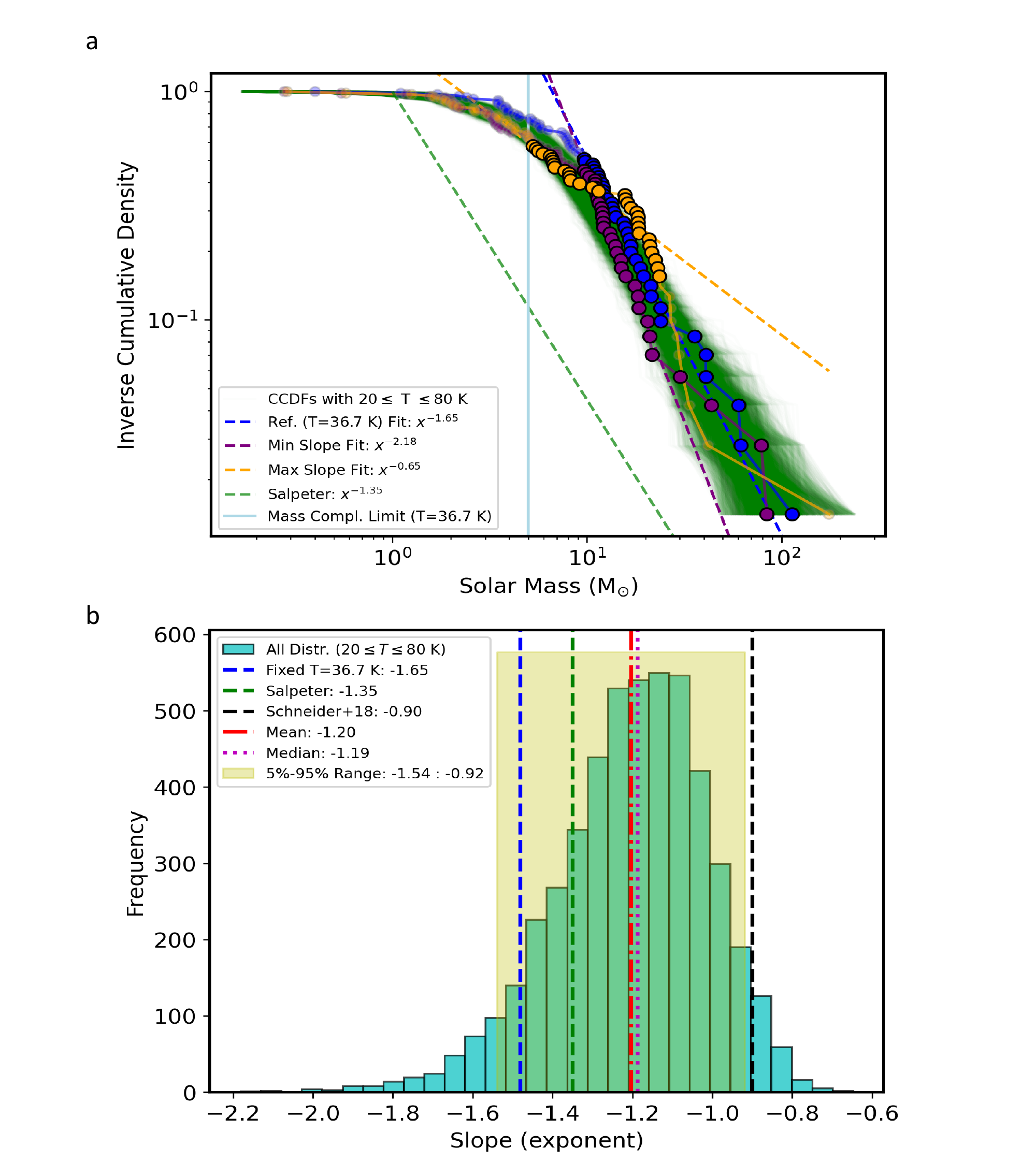}
\caption{\textbf{Distribution of the CMFs realization and their power-law exponents.} \textbf{a} 5000 Monte Carlo realizations of the inverse cumulative mass distributions of the 71 cores identified in our 4 proto-clusters. Each CMF distribution (light green) corresponds to one Monte Carlo realization of masses estimated using a temperature between $20\leq$T$\leq80$ K. The blue curve is the CMF distribution built using the reference temperature of 36.7 K. The yellow curve is the CMF distribution with the lowest value of the slope estimated in the overall distribution. The purple curve is the CMF distribution with the highest value of the slope. The semi-transparent points in the blue, yellow and purple distributions are the mass values not included in the power-law fit, all the other points are the ones used to estimate the power-law fit. The blue, yellow and purple dashed lines are the the power-law fit to the good points in these three reference curves. The green-dotted line is the Salpeter slope, showed for reference. The light blue vertical line represents the mass completeness for the blue distribution, estimated at T=36.7 K. Source data are provided as a Source Data file. \textbf{b} Histogram of the distribution of the power-law exponents of our 5000 Monte Carlo realization. The red vertical dash-dotted line is the mean value of the distribution, and the purple-dotted line is the median. The blue line is the slope exponent derived for our reference temperature CMF distribution. The green dashed line is the Salpeter slope value, and the dark blue line the IMF slope value dervied for 30Dor. The yellow shaded area represents the 5\%-95\% range of distribution of all our derived CMF slopes.}
\label{fig:CCMF}
\end{figure*}

\noindent \textbf{The CMF in 30Dor-10.}
The 5000 Monte Carlo realizations of the CMF are shown in Figure \ref{fig:CCMF} panel a as an inverse cumulative density function, and the distribution of the power-law tail exponents in Figure \ref{fig:CCMF} panel b. Each CMF exhibits two main regimes: flattening at low masses, potentially resembling a log-normal shape which is in principle broadly consistent with expectations for turbulence-dominated fragmentation \citep{Hennebelle08}, although high-resolution observations suggest that stars preferentially form within kinematically coherent filaments or fibers \citep{Hacar23}, and a power-law tail at higher masses, where gravitational collapse and competitive accretion dominate \citep{Motte98, Konyves15}. In some cases, a third regime appears, where the most massive objects follow a distinct slope, likely due to the limited number of sources in the final bins. Similar features are observed in the CMFs of ALMAGAL \citep{Coletta25} and ALMA-IMF \citep{Louvet24}. This study focuses on the well-defined power-law tail and its exponent, which we obtain by performing multiple power-law fits for each CMF, varying the starting point, ending point, and number of points within the defined limits (see the Methods section). The Kolmogorov-Smirnov (KS) distance was used to determine the best fit, i.e., the one with the minimum KS distance \citep{Louvet24}. This process yielded 5000 power-law fits, each with a different combination of starting mass (above the 30\textsuperscript{th} entry), ending mass (up to the highest value), and number of points used $(\geq30)$.

The most extreme power-law fits are shown in purple and yellow in Figure \ref{fig:CCMF} (upper panel), while the full distribution of exponents is presented in the histogram (lower panel). The mean and median slopes are $-1.20$ and $-1.19$, respectively, with a 5\%-95\% percentile range of $-1.54$ to $-0.92$. The CMF and best-fitting power-law for masses calculated using our reference temperature are highlighted in blue $(-1.65)$. The slope derived using a uniform core temperature falls outside the 5\%-95\% range, suggesting that this assumption may be overly restrictive, as temperature variations are expected within each proto-cluster. The Salpeter slope $(-1.35)$ is shown in green. The IMF exponent in 30Dor \citep[$-0.90$,][]{Schneider18}) also lies outside the 5\%-95\% range. Overall, the derived CMFs align well with the Salpeter slope, whereas the IMF slope in 30Dor remains an outlier.

To evaluate whether the Salpeter and 30Dor IMF slopes are consistent with the overall CMF exponent distribution, we computed the empirical $p$-value, accounting for the non-Gaussian and skewed nature of the distribution (Figure \ref{fig:CCMF}, lower panel). In such cases, the empirical $p$-value measures the probability of obtaining a value as extreme as the observed one by considering both tails of the distribution. It is determined by calculating the fraction of data points that are as extreme or more extreme than the observed value in both tails. A $p$-value $\leq 0.05$ indicates a highly unlikely result, while a $p$-value $>0.1$ suggests consistency with the distribution.

We find a $p$-value of 0.41 for the Salpeter slope and 0.08 for the IMF slope in 30Dor. The latter suggests evidence against the null hypothesis, indicating that a CMF power-law tail slope of $-0.90$ is highly unlikely. For completeness, we obtain a $p$-value of 0.04 for the slope derived using the reference temperature for all cores.

\section*{Discussion}\label{sec:conclusions}
Although the identification and characterization of cores in interferometric imaging are subject to large uncertainties, we have performed a careful core extraction in our ALMA images and derived 5000 realization of the power-law exponents of the global CMF in 30Dor-10 and the results indicate that the slope of the CMF, derived from the well-defined region of the slope distribution for the cores within the two most massive clumps of the 30Dor-10 region, is likely to follow a Salpeter-like slope. This result is in very good agreement with the CMF measured at 0.1 pc resolution in 30Dor-10 using $^{12}$CO data \citep{Indebetouw20}.

These results suggest that the formation process is multi-scale, with the 0.1 pc structures being directly linked to the cores formed at 2000 au scales. This link at different scales is predicted in theories of globally collapsing star-forming regions \citep{Vazquez-Semadeni25} and in simulations of gravo-turbulent collapse \citep{Nunez-Castineyra24}. As demonstrated before, our core mass distribution is also compatible with the formation of a canonical IMF \citep{Kroupa13, Yan23}, under the assumption that the observed IMF is directly inherited by the observed CMF with a defined core-to-star formation efficiency. But the observed slope of the IMF in 30Dor is somewhat shallower \citep{Schneider18, Schneider18b}. The presence of many main sequence massive stars could be interpreted as a fact that the stellar mass distribution does not originate in a simple way from the actual CMF. Instead, as suggested by multi-scale models of star formation, particularly in massive regions once gravity becomes dominant \citep{Nunez-Castineyra24}, and supported by several studies of the CMF in the Milky Way, the CMF evolves over time, transitioning from a Salpeter-like (or even steeper) distribution to a shallower one \citep{Pouteau23, Nony23, Armante24, Coletta25, Morii26}. Notably, these works represent the first efforts to separately characterize the pre-stellar and protostellar populations within the core samples, providing strong insight into the evolutionary trends of the CMF. In particular, in the W43 region, the CMF slope is steeper than Salpeter in the early, pre-stellar phase (with a value of $-1.46$), and becomes much shallower in the protostellar phase ($-0.64$), where star formation activity dominates \citep{Nony23}. A similar trend is observed in W33, where the slope evolves from $-1.69$ in the pre-stellar phase to $-1.04$ in the protostellar phase \citep{Armante24}. Same results have also been found in the analysis of the prestellar and protostellar cores identified in the ALMA Survey of 70 µm Dark High-mass Clumps in Early Stages (ASHES, \citep{Morii26}). These studies also indicate that the global CMF slope (the only one that we can derive in our study) reflects the dominant evolutionary stage of the population, likely protostellar-dominated in W43 (with a global CMF slope of $-0.96$), and pre-stellar dominated in W33-main (with a global CMF slope $-1.44$).

Based on the analysis of free-free contamination and the distribution of the slopes in our proto-clusters, it is likely that the observed regions in 30Dor-10 are all in a relatively early stage of evolution, with free-free emission contributing only about $5\%$ to the observed millimeter flux (see the Methods section). This result suggests that the observed global CMF could evolve towards a shallower slope over time, potentially reconciling the CMF slope with the IMF slope measured in 30Dor \citep{Schneider18, Schneider18b}. If this is the case, more evolved massive clumps in the 30Dor region could already show a shallower IMF, which would support a multi-epoch formation of stars in this region, as already suggested from observations of several different generations of stars \citep{Selman99, Crowther16, Rahner18}.

As detailed in Methods, we have also considered the possibility that our observed 2000 au cores may in fact consist of high-order multiple systems, as reported in several Galactic star-forming regions \citep{Li24}. Such multiplicity could, in principle, imply that the discrepancy between our derived CMF slope and the top-heavy IMF results from a resolution effect rather than an evolutionary trend. The analysis showed in the Methods Section, however, does not support this interpretation. The results clearly disfavor multiplicity as the dominant driver of the mismatch, reinforcing instead the conclusion that the most plausible explanation for the difference between the measured CMF slope and the observed IMF is the evolutionary stage of the clumps.

This study represents the attempt to examine continuum millimeter emission from star-forming regions in an external galaxy at about 2000 au resolution, an observational approach that has become routine for Milky Way proto-clusters thanks to facilities such as ALMA \citep{Motte22, Louvet24, Molinari25} and NOEMA \citep{Beuther18}. The findings presented here offer unique insights into star-formation processes in an environment significantly different from that of the Milky Way, demonstrating the feasibility of studies previously thought to be achievable only for local clusters. The investigation of these mechanisms across a diverse range of galaxies will become increasingly accessible with future facilities, enabling a systematic exploration of core populations across different environments: the key step toward unveiling the universal processes that govern star formation.


\section*{Methods}
Our data were taken as part of the ALMA project 2022.1.00917.S (PI: A. Traficante) in July 2023. The observations were centered on three massive clumps first identified in the ALMA Cycle 7 project 2019.1.00843.S (PI: R. Indebetouw) \citep{Indebetouw13} with a resolution of approximately 0.5\,pc. Here we employ their nomenclature, and refer to them as Clump 4, Clump 52 and Clump 72. These clumps are among the most massive in the region, with CO masses of about 2200 \Msun, 8500 \Msun\ and 7400 \Msun\ within a beam-deconvolved radius of about 0.3 pc, 0.5 pc and 0.5 pc for clump 4, 52 and 72 respectively \citep[][]{Indebetouw13}. 

Each clump was mapped in the C43-8 configuration, covering a $0.35\arcmin\times 0.35\arcmin$ squared region. At the distance of the LMC this corresponds to a region of approximately $5\ \mathrm{pc}\times5$\,pc. The maximum recoverable scale of our configuration is about 0.8 pc, sufficient to encompass the full extent of the identified clumps. This approach ensures the recovery of all significant emission associated with the clumps while filtering out the most extended emission components through the interferometer (whose contribution does not dominate as demonstrated later in this section), thus resulting in maps characterized by a relatively uniform noise level. The synthesized beam is $0.048\arcsec\times0.038\arcsec$ in all clumps, which corresponds to a linear resolution of about $2400\times\ 1900$ au. The scientific data comes from the ALMA pipeline reduction vs. 2023.1.0.124 \citep{Hunter23} and using CASA 6.5.4.9 \citep{CASA22}. The continuum maps have been obtained by combining the line-free channels of the four spectral windows required in our observational setup which have the required sensitivity of about $10\ \mu$Jy/beam.

\subsection*{Source extraction and analysis}\label{sec:analysis}
The flux properties of the hosting clumps and of the identified ALMA objects have been extracted with \Hypy, a Python implementation of the original IDL version of \Hyp\ \citep{Traficante15a}, which preserves its full functionality while introducing several key improvements that enhance its performance and flexibility. This code is designed to deal with complex background, highly blended sources and multi-wavelength extraction, and proved to be well suited for source extraction and photometry in ALMA images \citep{Traficante23, Coletta25}. It first fits a 2D-Gaussian for each identified peak together with a background and secondly performs an aperture photometry within the background subtracted region defined by the 2 FWHMs (used as aperture radii) and the position angle of each source. 

Further information about the extraction and identification of the final set of 71 cores performed with \Hypy  are given in SI. Photometric and physical properties of these 71 cores can be found in Supplementary Data 1. Flux completeness limit of our extraction estimated by injecting synthetic sources in the real maps are also discussed in Supplementary Discussion, Supplementary Table 1 and Supplementary Figure 7.

\subsection*{Potential artifacts from interferometric imaging}\label{app:interferometric_artifacts}
We have examined a potential bias intrinsic to high-resolution interferometric imaging: the spurious generation of apparent substructures due to incomplete sampling of spatially smooth emission. Such imaging artifacts can arise when large-scale emission is filtered out in the Fourier domain, a limitation of interferometric observations that lack adequate short-spacing information, as observed in particular in low-mass star-forming regions \citep[e.g.,][]{Caselli19, Tokuda20}.  The details of the analysis are given in Supplementary Discussion and Supplementary Figures 2-4. From this analysis, we conclude that the compact sources identified in our high-resolution images are unlikely to be spurious, and the impact of missing flux due to incomplete \textit{uv} coverage on our analysis is negligible.

\subsection*{Reliability of the core identification}\label{app:cores_reliability}
The incomplete sampling of spatially smooth emission may not be the only mechanism capable of producing spurious artifacts in interferometric images. In particular, gaps in the \textit{uv} coverage and the contribution of the sidelobes from the synthesized beam can also introduce weak, compact features within diffuse emission, which may subsequently be identified as real cores by any extraction algorithms. A rigorous assessment of this effect requires the generation of realistic input models for the smooth background emission, followed by full interferometric simulations that accurately reproduce our ALMA observational setup. To this end, we exploited the Rosetta Stone (RS) framework \citep{Lebreuilly25, Tung25, Nucara25}, which provides the most complete end-to-end pipeline for converting numerical simulations into synthetic ALMA observations, incorporating with high fidelity the actual array configuration, scheduling blocks, source elevations, and all other observational parameters. This approach allows for an almost one-to-one comparison between synthetic and real interferometric images \citep{Nucara25}. Although the correct identification of cores and core boundaries in interferometric imaging remains a challenge, all our tests (detailed in Supplementary Discussion, see also Supplementary Figure 5) support that the relatively low and smooth background associated with our intrinsically weak sources is not able to generate false peaks above our $4\sigma$ threshold and confirms that our adopted $4\sigma$ threshold is reliable: all sources detected in the real map, including those identified against the more diffuse emission, are highly likely to correspond to genuine compact structures.

We further examine how the estimation of the power-law exponent of our CMF would vary under the conservative assumption that a fraction of the weakest detected sources may be spurious. Even this stress test confirms that our inferred slopes remain stable and fully consistent with a Salpeter-like distribution, rather than with a top-heavy one.

\subsection*{Assessing core gravitational boundedness}\label{app:gravitational_boundedness}
An important requirement for building a meaningful CMF is to ensure that the identified structures are not transient or unbound fluctuations, but rather real gravitationally bound cores. This analysis is detailed in Supplementary Discussion and suggests that the vast majority of our cores, particularly those used to derive the slope of our CMF, and more generally the 65 out of 71 sources with radii $\geq 2500$ au are likely gravitationally bound. As showed in Supplementary Figure 6, the minimum non-thermal velocity dispersion required to unbind them exceeds, in most cases significantly, the values typically observed in Galactic dense cores of similar size in some of the most active high-mass star-forming regions of the Milky Way.

\subsection*{Free-free contamination in the 30Dor-10 clusters}\label{sec:free_free_estim}
The integrated flux $F_{\mathrm{INT}}$ measured with \Hypy\ for each source can be used to estimate the masses of a core $M_{\mathrm{c}}$ under some assumptions. Before converting the observed flux into mass, the first crucial step is to verify that the measured mm flux originates from thermal emission of the cold dust envelope and is not significantly contaminated by the tail of radio free-free emission.

To assess potential free-free contamination, we conducted an extensive analysis by combining H$_{\alpha}$ emission line data from the HST and Pa$_{\alpha}$ emission line data from the JWST of the 30Dor-10 region. In summary, we first estimated the extinction-corrected H$_{\alpha}^{corr}$ emission from HST by combining it with the Pa$_{\alpha}$ emission observed with JWST. We then re-scaled H$_{\alpha}^{corr}$ to 1.3 mm to estimate the contribution of free-free emission to our ALMA flux. The step-by-step procedure of our pipeline is described in Supplementary Discussion. In Supplementary Figure 8 we show the maps of the derived free-free contamination at 1.3 mm for our three clumps, and in Supplementary Table 2 we summarize our findings. This analysis confirms that free-free contamination remains low $(\leq5\%-6\%)$ in both clump 52 and 72 (and it is likely above 30\% in clump 4). Consequently, the presence of nearby HII regions does not significantly impact the dust emission observed at 1.3 mm in these clumps. Additionally, the minimal free-free contribution to the mm flux suggests that these proto-clusters are still in an early evolutionary stage, with no indication of significant HII regions having yet developed within them.

\subsection*{Physical properties of the extracted cores}\label{app:cores_props}
The physical properties of each core were derived from the measured fluxes using the standard greybody emission formula, assuming that dust is optically thin in our regime \citep{Sanhueza19, Svoboda19, Coletta25}. The calculations adopt the reference values of distance $d= 50$ kpc, gas-to-dust ratio $R_{gd}=500$, dust absorption coefficient $\kappa_{1.3\mathrm{mm}} = 1.16$ cm$^2$ g$^{-1}$ and dust temperature $T=36.7$ K. These values were selected based on the following considerations: the adopted distance of 50 kpc is consistent with the measurement of $49.60\pm0.63$ kpc \citep{Pietrzynski19}. The gas-to-dust ratio $R_{gd}$ varies as function of metallicity and gas surface density \citep{Roman-Duval17}. In the LMC, $R_{gd}$ ranges from 1500 to 500 in regions with gas surface densities of $10–100$ M$_{\odot}$ pc$^{-2}$, which is significantly higher than the standard value of 100 measured in the Milky Way. This is likely due to the lower metallicity of the LMC, making it more comparable to the values observed in high-redshift dwarf galaxies \citep{Curti20}. Spectroscopic studies of Red Giant Branch stars in the LMC indicate an average metallicity of about -0.38 dex with a shallow gradient across the disk \citep{Choudhury16}, although gas-phase metallicity maps reveal a more complex structure \citep{Lah24}. We have decided to adopt $R_{gd}=500$, the most appropriate value for high-mass star-forming regions in the LMC \citep{Brunetti19}. The dust absorption coefficient $\kappa_\nu$ depends on grain composition and may vary across different lines of sight \citep{Ossenkopf94}. Extensive opacity mapping provides estimates of $\kappa_\nu$ in the LMC \citep{Gordon14}, and for this study we use $\kappa_{1.3\mathrm{mm}} = 1.16$ cm$^2$ g$^{-1}$, derived by averaging opacity values from 30Dor, N159W, and N159E \citep{Brunetti19}.

Different assumptions on the distance $d$, the gas-to-dust ratio $R_{\rm gd}$, or the dust opacity $\kappa_\nu$ affect the absolute core masses, but do not impact the slope of the CMF, provided that these parameters are approximately uniform across the sample. In the case of 30Dor-10, the cores lie within a limited region of the LMC and the internal variation in $R_{\rm gd}$ and $\kappa_\nu$ is expected to be modest. Observational studies suggest that $R_{\rm gd}$ in the LMC varies by less than a factor of two across different regions \citep[][]{Galliano11, Roman-Duval14}, and variations in $\kappa_\nu$ at millimeter wavelengths are typically limited to a factor of 1.5 within a given molecular cloud \citep{Ossenkopf94, Gordon14}. We have therefore adopted fixed values for $R_{\rm gd}$ and $\kappa_\nu$, consistent with previous studies, under the assumption that spatial variations within our sample are not sufficient to bias the derived CMF slope.

Supplementary Data 1 contains the photometric and physical properties of the 71 cores extracted with \Hypy\ and associated with the 4 proto-clusters.

\subsection*{Estimation of the CMF power-law exponent}
To obtain a well-constrained estimate of the CMF power-law exponent $\Gamma$ from the formula $d\mathrm{N}/d\mathrm{Log(m)}\propto m^{-\Gamma}$, we systematically examined three key parameters: the starting value of the fit, the ending value of the fit and the number of points used to perform the fit. To fully vo validate the results of the CMF analysis we have yielded 5000 power-law fits, which were then used to derive the probability distribution of the power-law exponents. We have also further examined the sensitivity of the power-law exponent distribution to the number of Monte Carlo realizations. The details of this analysis are in Supplementary Table 3.

\subsection*{Variation of the CMF exponent as function of the selected cores}
Although we remain confident in the reliability of our core identification, we have nonetheless explored the possibility that a subset of the detected cores, particularly those with the lowest contrast relative to the local background (i.e. those with the lowest peak S/N), may be spurious. These tests aim to further explore the possibility that interferometric imaging may produce spurious compact sources, thereby affecting the reliability of the primary catalog. This analysis shows that interferometric imaging could produce spurious sources, in particular objects with a relatively low contrast compared to the background. However, by selecting a sub-sample of cores with high contrast relative to the local background, we focus exclusively on the most reliable detections, while simultaneously reducing ambiguities in the core extraction and definition process. As detailed in Supplementary Table 4, these tests provide additional support for our findings and further strengthen the conclusion that the observed CMF is consistent with a Salpeter-like slope.

\subsection*{Variation of the CMF exponent as function of the core multiplicity}
As shown in massive star-forming regions in our Galaxy, binary systems and, in some cases, higher-order multiples are expected to arise within massive cores observed at spatial scales of 2000 au, with individual components separated by about 150–700 au \citep{Li24}. Consequently, the exponent of the CMF measured at these sub-$2000$ au scales may differ substantially from the CMF slope inferred at 2000 au. The empirical correlation observed between the CMF slope at 0.1 pc and that at 2000 au may therefore break down at smaller radii, implying that the discrepancy between the IMF slope and the CMF slope on 2000 au scales could, in principle, be driven by resolution effects rather than by evolutionary effects. The results are detailed in Supplementary Tables 5-6 and they strongly support the conclusion that, at the evolutionary stage probed by our observations, the CMF remains fundamentally incompatible with the IMF and that unresolved multiplicity alone cannot reconcile the observed discrepancy between the two slopes.

\section*{Data availability}
All Tables and images generated during and/or analysed during the current study are available in the public GitHub repository \texttt{https://github.com/Alessio-Traficante/LMC\_30Dor10\_metadata}. Supplementary Data 1 are also available at the same GitHub repository. All other relevant data are available from the corresponding author on request.

\section*{Code availability}
The \Hypy\ code is publicly available at the GitHub repository \texttt{https://github.com/Alessio-Traficante/hyper-py} and at the Zenodo repository \texttt{https://zenodo.org/records/18469455}. The python codes used to derive the results have been deposited in a GitHub repository available at \texttt{https://github.com/Alessio-Traficante/LMC\_30Dor10\_metadata}.



\bibliography{sn-bibliography}

\bmhead{Acknowledgements}
This paper makes use of the following ALMA data: 2022.1.00917.S. ALMA is a partnership of ESO (representing its member states), NSF (USA) and NINS (Japan), together with NRC (Canada), MOST and ASIAA (Taiwan), and KASI (Republic of Korea), in cooperation with the Republic of Chile. The Joint ALMA Observatory is operated by ESO, AUI/NRAO and NAOJ. R.I. was partially supported during this work by NSF award 2009624 to the University of Virginia. The National Radio Astronomy Observatory and Green Bank Observatory are facilities of the U.S. National Science Foundation operated under cooperative agreement by Associated Universities, Inc. A.T., R.S.K. and S.M. acknowledge financial support from the ERC via Synergy Grant ``ECOGAL'' (project ID 855130). R.S.K. acknowledges financial support from the German Excellence Strategy via the Heidelberg Cluster ``STRUCTURES'' (EXC 2181 - 390900948), and from the German Ministry for Economic Affairs and Climate Action in project ``MAINN'' (funding ID 50OO2206).


\section*{Author contributions}
A.T. performed the source extraction and data analysis. MJD helped to generate the images in Figure 2. RI provided combined data with low-resolution, ancillary ALMA data and performed the analysis on the combined dataset. TW provided the JWST images of the field. E.S. provided the HST images of the field. A.T. directed the project and wrote the manuscript. All authors commented on the manuscript. 

\section*{Competing interests}
The authors declare no competing interests.

\end{document}